\documentstyle[12pt,newpasp,twoside,epsf]{article}

\def\edcomment#1{\iffalse\marginpar{\raggedright\sl#1\/}\else\relax\fi}
\reversemarginpar

\begin{document}

\title{Panel Discussion on Observing Simulations and Simulating Observations}

\author{Piet Hut\altaffilmark{1},
	Adrienne Cool\altaffilmark{2},
	Charles Bailyn\altaffilmark{3},
	Steve McMillan\altaffilmark{4},
	Mario Livio\altaffilmark{5},
	Mike Shara\altaffilmark{6},
    	}
\altaffiltext{1}{Institute for Advanced Study,
		 Princeton, NJ 08540, USA}
\altaffiltext{2}{Department of Physics and Astrophysics,
		 San Francisco State University,
		 San Francisco, CA 94132, USA}
\altaffiltext{3}{Department of Astronomy,
		 Yale University,
		 New Haven, CT 06520, USA}
\altaffiltext{4}{Department of Physics,
		 Drexel University,
                 Philadelphia, PA 19104, USA}
\altaffiltext{5}{Space Telescope Science Institute,
		 3700 San Martin Drive,
		 Baltimore, MD 21218, USA}
\altaffiltext{6}{Department of Astrophysics,
		 American Museum of Natural History,
		 Central Park West at 79th St.,
		 New York, NY 10024, USA
		 }

\begin{abstract}
N-body simulations of star cluster evolution have reached a high
degree of realism, by incorporating more and more elements of stellar
dynamics, stellar evolution, and hydrodynamics.  At the end of this
conference, six participants discussed how to present the increasingly
realistic data from star cluster simulations in a way that is most
useful for a direct comparison with observations.
\end{abstract}

\section{Introduction, by Piet Hut}

Only after organizing this panel did I realize that the acronym of the
title, OSSO, has the meaning of `bone' in Italian -- perhaps fitting
for a conference in the American Museum of Natural History, which is
generally associated with paleontology.  Also, until recently, star
cluster simulations have been rather skeletal in that they have left
out much of the essential physics, including stellar evolution.  Only
recently have we begun to put flesh and skin on the bones of our
simulations, as we have seen during the conference.

     Two stumbling blocks have prevented a production of realistic
simulations.  The first one is related to a lack of hardware speed.
Although Fokker-Planck and gas models have provided important insight
into the dynamics of star clusters, they are not well suited for
studying the dynamical effects of a significant binary population.
Therefore, direct $N$-body simulations are called for.  On a typical
workstation, with a speed of order of a Gflops, one can now run a
thousand-body run overnight, and with more patience a 5,000-body run
can be performed if one is willing to wait a month or more.  To model
the richest open clusters, with 50,000 stars, a Tflops speed is called
for, and modeling a typical globular cluster, with at least 500,000
stars, requires a Pflops speed.  In 1995, the Tflops barrier was broken
with the GRAPE-4, and the GRAPE-6 is expected to reach a speed of
100 Tflops in 2001.  Finally, the GRAPE-8 is expected to deliver
several Pflops well before the end of the decade, thus allowing the
modeling of any globular cluster.

     However, with hardware solutions being just around the corner,
software limitations are making themselves felt all over the place --
much like everywhere else in the world nowadays.  For one thing, what
is desperately needed is access to simple stellar evolution codes that
are robust enough to serve as modules in star cluster simulation codes.
It would be ideal to construct models of blue stragglers on the fly,
immediately after they form in a collision, and then follow their
specific evolution, without having to interpolate between tracks based
on the very different evolution that starts with zero-age main sequence
stars.  It seems to be a well-kept secret that after almost a half
century of numerical stellar evolution work, still no code can follow
the full evolution of a single star without human intervention,
something that is impractical once we are dealing with hundreds of
thousands of stars.

     Other software challenges involve the visualization of the Tbytes
that are currently generated with our Tflops computations, and the
Pbytes that will be generated towards the end of the decade.
Constructing simulation archives, with efficient ways to interrogate
the data and to pipe relevant data subsets to other geographical
locations, are tasks that we are only beginning to confront.  And in
order to make contact with the observations, the most direct way will
be to simulate observations of the simulations -- a software S.O.S.
reaction to the coming data flood.

\section{Observing Simulations, by Adrienne Cool}

Simon, Piet, Steve, and Jun have taken the initiative to try to bridge
the gap that sometimes exists between observers and theorists working
on star clusters.  This is a very welcome development, since while we
may not always like to admit it, it can be surprisingly hard to find
solid points of contact.  Questions as apparently trivial as ``what is
the core radius of this cluster?'' turn out, as became clear at the
workshop, to involve numerous subtleties that often get swept under
the rug.

It's time to improve this situation.  What with the rapid advances
in the theoretical modeling of clusters, and the richness of the
observational data being collected almost daily, increasingly direct
and meaningful comparisons between theory and observation are
beginning to be possible.

So how to begin to bridge the gap?  The approach that Simon and
collaborators have taken involves ``observing simulations.''  They are
collecting the results of their cluster simulations, generating from
them simulated observations, and offering these up to observers for
analysis.  The ways in which this sort of approach can be useful are
just beginning to be explored.  One obvious utility is to see how
closely what you get out resembles what you put in.  How accurately is
the main-sequence luminosity function, the main sequence binary
fraction, or the white dwarf cooling sequence reproduced?  Do the core
radius and tidal radius extracted from the simulated data set match
the core radius and tidal radius of the simulated star cluster from
whence it came?

To some extent, questions like this can be (and have been) addressed
by widely used ``artificial star tests.''  One complication, as any
observer will instantly point out, is that how much you can extract
from the data will depend sensitively on its nature and quality.  What
you choose to simulate by way of filters, pixel size, psf structure,
cosmic rays, artifacts (the list goes on and on) will all have an
impact on the results.  Exploring such questions could be interesting
from the point of view of using simulations to test the feasibility of
various kinds of measurements, but is perhaps beside the point here.
In the present context, the more important opportunity that this next
generation of artificial star tests affords is to find out whether
observers and theorists are even speaking the same language.  In some
cases (core radius is a good example) we already know that we aren't.
Communicating through simulated observations can provide the impetus
to find a common language.

Perhaps even more interesting is to consider how to take advantage of
what is really new: the fact that there is a fully 3-dimensional
cluster to work with.  This means you can ``observe'' the cluster from
an arbitrary point in space.  In principle, you could also observe the
cluster at a variety of equivalent (or not) points in time.  Thus,
comparisons can be made between results obtained by observing the very
same cluster from different places or times.  Beyond the inherent
appeal of even the imagined freedom to move about in space and time,
this approach could help address questions related to small number
statistics, and provide the means to explore potentially subtle
projection effects.

Taking this a step further, Mike Shara has challenged us all to think
about what can be learned not just from analyzing simulated
observations of a cluster frozen at a particular moment in time, but
with rendered 3-dimensional dynamic simulations, like the one we all
got a taste of at the opening reception.  This kind of viewing can be
done in real time, in the sense that one can make choices about where
to move and what to look at on the fly (so to speak).  Opportunities to
observe simulations in this way could this be a boost to developing
intuition about cluster dynamics.  Observing simulations and
discussing them in small groups could also provide an intriguing new
forum for enhancing communication between theorists and observers.

\section{Defining Definitions, by Charles Bailyn}

There is often confusion regarding the meaning of a number of commonly
used terms relating to the dynamics of clusters.  Observers and
theoreticians use these terms without defining them, and in ways which
make it difficult to compare theoretical and observational results.
In some cases it is not clear what the appropriate definition ought
to be.  We feel it is important for all workers to provide careful
definitions when they use these terms, and to make an effort to record
results in ways that are not merely clear, but useful to the widest
possible audience.

The chief offender seems to be "core radius".  This is a well-defined
parameter of a King model, but it is not clear how it should be defined
in situations in which a King model cannot be fit, either because the
data are not extensive enough or because the distribution is poorly
fit by a King model.  It is often assumed that $r_c$ represents a
distance at which the density falls to half the central density.
(Note that the conversion of density to projected density is a continuing
difficulty in comparing observations of real clusters and observations
of simulations --- it is MUCH better to project models into observational
space, so cluster simulators are strongly encouraged to quote projected
densities).  But this definition of $r_c$ begs the question of how a
central value of the density is defined, either observationally or
theoretically.  Not only is a ``central'' density an instrumentally
defined term for observers, but it is undefined theoretically too, due to
the stochastic nature of the inner regions of clusters.

One suggestion would be to define a ratio of radii, and a ratio of
encircled densities, and scale the radii until the density ratio is
correct.  The density within the inner radius might then provide a
robust measure of the ``central'' density, while the outer radius
would define a ``core radius''.  But it is not clear what the appropriate
ratios would be, or even whether such a definition would in fact
be robust, stable, or repeatable.  Clearly, further investigation is
required to create appropriate definitions for observers of simulated or
real data.

Other terms which create confusion include ``collapsed core'', ``tidal
radius'', ``half-mass radius'' and ``primordial binary''.  A collapsed
core is generally considered to be one whose density rises continuously
to the center.  But the problems of defining central density arise here
in particularly virulent form.  It is notable that the core of 47 Tuc,
perhaps the best studied globular cluster, is sometimes described as
collapsed, and sometimes not.  Tidal radius is another term which is
well-defined in the context of a King model, but it is NOT synonymous with
the largest distance a cluster member can be from the cluster center.
This difference is crucial to remember when computing or observing the
half-mass radius, since potentially significant cluster members can lie
beyond the tidal radius.  Finally, there was general agreement that
primordial binaries refer to binaries in which the two component were
bound at the start of the cluster lifetime, even if the parameters of
the binary orbit have been significantly altered by subsequent dynamical
interactions.

\section{Simulation Bottlenecks, by Steve McMillan}

The impending appearance of GRAPE-6 will significantly lower the main
computational barrier to direct $N$-body simulations of star clusters
and dense stellar systems.  From a purely computational standpoint at
least, with existing software and a reasonably fast host, we can say
loosely that GRAPE-4 enables the study of many, but not all, open
clusters, while GRAPE-6 will allow us to perform simulations of all
open clusters and at least the smaller globulars.  It then becomes
feasible to contemplate ``throughput'' experiments, in which one
systematically varies the assumed cluster initial conditions and
compares the results directly with real systems observed today.
However, the old adage ``garbage in, garbage out'' continues to apply.
The sources of ``garbage'' in this case are uncertainties in (i) the
initial models ($t=0$), (ii) evolutionary processes ($0<t<\mbox{now}$),
and (iii) interpretation of observations ($t=\mbox{now}$).

The initial state of a cluster is not well known.  Perhaps the most
important uncertainty, from both the dynamical and the observational
standpoint, stems from the properties of the primordial binary
population: numbers, masses, mass ratios, and periods.  It is standard
practice to assume an initially homogeneous sample (i.e.~no initial
mass segregation), but there seems no particular reason to suppose
that this is really the case, and in fact there are arguments to
suggest the opposite---that more massive stars and binaries will form
preferentially in the denser central regions.  Finally, the question
of what exactly is meant by ``$t=0$'' is also unresolved.  Low-mass
stars may take hundreds of millions or even billions of years to reach
the main sequence and, at the time of formation, may have radii
hundreds of times greater than typically assumed in the models.
Whether or not this significantly affects cluster evolution remains to
be seen, but it at least highlights the fact that very substantial
uncertainties exist in the initial models.

There are many open questions concerning the essential physics.
The proper treatment of binary and stellar evolution is critical
if we wish to interpret cluster observations in the light of model
simulations.  While stellar evolution theory is sufficiently advanced
that the evolution of most stars can be modeled by interpolation
between standard tracks, even here there are areas of uncertainty.
Specifically, the evolutionary tracks of high-mass stars, and
especially merger products, remain poorly determined.  The largest
uncertainties are again associated with binaries, and some phases of
binary evolution are currently modeled in a very heuristic fashion.
In particular, the lifetimes of contact binaries and their descendents
are largely unknown.

Finally, there are questions concerning the interpretation of
observations.  Excellent cluster data are now becoming available, both
for open clusters (e.g.~the WIYN Open Cluster Survey) and for globulars
(with high-precision HST and Chandra studies now commonplace).  However,
Much data analysis still involves the use of unrealistically simple
dynamical templates.  Obvious examples are the use of multi-mass King
models, or dynamical models that neglect binaries, stellar evolution,
and/or the influence of Galactic tides, as standards against which
observations are gauged.  For the foreseeable future, the preferred
approach to making the comparisons will be to project the simulations
into the observational plane, using simulated telescope, filter, and
detector characteristics as appropriate, with extra field stars and
obscuration if desired, and to ``observe'' these model systems using
the same techniques as would be applied to real clusters.

\section{Mergers in the Universe, by Mario Livio}

I like to look at collisions from a more global point of view. For
example, we know from the Medium Deep Survey, and from the two Hubble
Deep Fields, that the fraction of faint galaxies with irregular
morphologies increases with redshift.  Many of these are suggestive of
merging or colliding systems.  Similarly, there is evidence for
galaxies being physically smaller in size beyond redshift z=1.  All of
these observational facts can be interpreted in terms of hierarchical
galaxy formation -- in the high redshift universe we are seeing the
"building blocks" of today's galaxies.

A related question is that of the formation of clusters in galaxy
collisions.  For example in the "Antennae" (NGC 4038/4039) merging
galaxies, HST has detected between 800 and 8000 luminous young
clusters that formed in the collision process. The luminosity function
of these clusters is (to first order) a power law, with an exponent of
-2.1. It would be interesting to see whether theoretical simulations
can reproduce such a power law.

Another question concerns the central supermassive black holes in
galaxies.  Such black holes appear now to reside at the center of
essentially all galaxies (and their mass is correlated with the
velocity dispersion).  One of the issues in active galactic nuclei is
whether these black holes grow mainly by accretion or through mergers.
The hierarchical galaxy formation picture may suggest that mergers
should be an important growth mechanism.

All of the above suggest that collisions are not only important for
the fate of individual stars, but also more globally.

\section{A Call for Predictions, by Mike Shara}

The computational power and astrophysical brain power being brought to
bear today on the structure, evolution and populations of dense star
clusters is nothing short of exhilarating.  There is some danger,
however, that we are becoming victims of our own success with problems
analogous to those being faced by observational astronomers: data
floods of epic proportions.

Observing the simulations is an ever increasing challenge in a
(soon-to-be) Petabyte world.  I predict that we will need more and
more collaboration from our computer scientist colleagues, and novel
display devices (like the Hayden Planetarium digital dome) to let our
eyes and brains pick out the data diamonds from the numerical gravel.
The astrophysics department at the American Museum of Natural History
has committed a significant amount of digital dome night time
(equivalent to ground based telescope dark time) to attacking the
visualization problem.  Ongoing collaborations of star cluster
simulators (particularly GRAPE aficionados), computer scientists and
Planetarium visualization experts are aimed at producing a 30 meter
"digital telescope" in the heart of Manhattan, long before CELT or OWL
come on line.

My cheers and challenges are directed at those intrepid
theorist/numericist astrophysicists who try to simulate observations
of dense star clusters.  A poster child for this difficult endeavour
is the gutsy paper of Di Stefano and Rappaport (ApJ 423, 274 (1994))
who boldly predicted the existence of about 100 cataclysmic variables
in each of Omega Cen and 47 Tuc.  Hubble Space Telescope surveys in
narrowband Halpha and for erupting dwarf novae have found a few
cataclysmics in several clusters surveyed so far, but nothing like the
100 predicted.  Much more sensitive Hubble and Chandra observations
now in hand may yet turn the tide.  My point, though, is that SPECIFIC
predictions of the numbers and types of unusual stellar species in
clusters are rare, but EXTREMELY valuable as tempting carrots to
observers.  Rarer still are those simulations marrying stellar
populations models with realistic dynamics.

An important goal of the coming generation of GRAPE-6 simulations
(coupled to evolution/population codes) should be specific predictions
of the numbers, observational characteristics and spatial distributions
of cataclysmic binaries, millisecond pulsars, neutron stars and red
giants in many of the globular clusters of the Milky Way.  This
will help observers push for larger allocations of telescope time
(particularly on HST and Chandra) to produce tougher constraints,
driving simulators to more sophisticated predictions. The populous
star clusters of the Magellanic Clouds should not be ignored in
this effort, as HST and Chandra can detect some of their stellar
exotica . . .  and the simulated observations can suggest how these
clusters differ from those of our Milky Way.

\end{document}